\documentclass[a4paper,fleqn,usenatbib]{mnras}
\usepackage{graphicx}

\title[Cool white dwarf companions to four millisecond pulsars]{Cool
  white dwarf companions to four millisecond pulsars}

\author[C. G. Bassa et al.]{C. G. Bassa$^{1}$\thanks{bassa@astron.nl},
  J. Antoniadis$^{2}$,
  F. Camilo$^{3}$,
  I. Cognard$^{4}$,
  D. Koester$^{5}$,
  \newauthor
  M. Kramer$^{6, 7}$,
  S. R. Ransom$^{8}$,
  B. W. Stappers$^{7}$
  \\
  $^1$ASTRON, the Netherlands Institute for Radio Astronomy, Postbus 2, 7900 AA, Dwingeloo, the Netherlands\\
  $^2$Dunlap Institute for Astronomy and Astrophysics, University of Toronto, 50 St.\ George Street, Toronto, Ontario M5S 3H4, Canada\\
  $^3$Columbia Astrophysics Laboratory, Columbia University, New York, NY 10027, USA\\
  $^4$Station de Radioastronomie de Nan\c cay, Observatoire de Paris, 18330 Nan\c cay, France\\
  $^5$Institut f\"ur Theoretische Physik und Astrophysik, University of Kiel, D-24098 Kiel, Germany\\
  $^6$Max Planck Institut f\"ur Radioastronomie, Auf dem H\"ugel 69, 53121 Bonn, Germany\\
  $^7$Jodrell Bank Centre for Astrophysics, The University of Manchester, Manchester, M13\,9PL, United Kingdom\\
  $^8$National Radio Astronomy Observatory, Charlottesville, VA 22903, USA\\
}

\date{Accepted 2015 November 2.  Received 2015 October 9; in original form 2015 August 28}

\pubyear{2015}

\begin{document}
\label{firstpage}
\pagerange{\pageref{firstpage}--\pageref{lastpage}}
\maketitle

\begin{abstract}
  We report on photometric and spectroscopic observations of white
  dwarf companions to four binary radio millisecond pulsars, leading
  to the discovery of companions to PSRs\,J0614$-$3329, J1231$-$1411
  and J2017+0603. We place limits on the brightness of the companion
  to PSR\,J0613$-$0200. Optical spectroscopy of the companion to
  PSR\,J0614$-$3329 identifies it as a DA type white dwarf with a
  temperature of $T_\mathrm{eff}=6460\pm80$\,K, a surface gravity
  $\log g=7.0\pm0.2$\,cgs and a mass of
  $M_\mathrm{WD}=0.24\pm0.04$\,M$_\odot$. We find that the distance to
  PSR\,J0614$-$3329 is smaller than previously estimated, removing the
  need for the pulsar to have an unrealistically high $\gamma$-ray
  efficiency. Comparing the photometry with predictions from white
  dwarf cooling models allows us to estimate temperatures and cooling
  ages of the companions to PSRs\,J0613$-$0200, J1231$-$1411 and
  J2017+0603. We find that the white dwarfs in these systems are cool
  $T_\mathrm{eff}<4000$\,K and old $\ga5$\,Gyr. Thin Hydrogen
  envelopes are required for these white dwarfs to cool to the
  observed temperatures, and we suggest that besides Hydrogen shell
  flashes, irradiation driven mass loss by the pulsar may have been
  important.
\end{abstract}


\section{Introduction}
Over the last few years the number of Galactic radio millisecond
pulsars has increased dramatically \citep{ran13}. This increase is due
to improved hardware used in all-sky pulsar searches at the Arecibo
\citep{cfl+06}, Green Bank \citep{hrk+08,blr+13}, Parkes
\citep{kjs+10} and Effelsberg \citep{bck+13} telescopes, as well as
targeted searches for pulsars associated with unidentified
$\gamma$-ray sources recently discovered by the \textit{Fermi}
$\gamma$-ray space telescope (see \citealt{rap+12} for a summary and
references).

\begin{table*}
  \footnotesize \centering
  \caption[]{Properties of the pulsars under investigation. Distances
    $d_\mathrm{DM}$ are estimated from the observed dispersion measure
    ($\mathrm{DM}$) and the Galactic extinction model of
    \citet{cl02}. Characteristic ages ($\tau_\mathrm{c}$) and spindown
    luminosities ($L_\mathrm{SD}$) are estimated from the observed
    spin period $P$ and the intrinsic spin period derivative
    $\dot{P}_\mathrm{int}$, which is the observed spin period
    derivative corrected for the \citet{shk70} effect. Minimum and
    median companion masses are estimated assuming a 1.4\,M$_\odot$
    pulsar and orbital inclinations of $i=90\degr$ for
    $m_\mathrm{c,min}$ and $i=60\degr$ for $m_\mathrm{c,med}$. The
    theoretical $P_\mathrm{b}$-$M_\mathrm{c}$ relation by
    \citet{lrp+11} is used for the predicted companion mass
    $M_\mathrm{c,pred}$. The $P\mathrm{b}$-$M_\mathrm{c}$ relation by
    \citet{ts99} predicts companion masses that are 0.016\,M$_\odot$
    heavier. The positional offsets $\Delta \alpha$ and $\Delta
    \delta$ denote the offset in position with respect to the
    pulsar. }\label{tab:parameters} \renewcommand{\footnoterule}{}
  \begin{tabular}{lllll}
    \hline \hline
    Parameter & PSR\,J0613$-$0200 & PSR\,J0614$-$3329 & PSR\,J1231$-$1411 & PSR\,J2017+0603 \\
    \hline
    Spin period  $P$ (ms) & 3.061844 & 3.148670 & 3.683879 & 2.896216 \\
    Period derivative $\dot{P}_\mathrm{int}$ & $9.590\times10^{-21}$ & $1.74\times10^{-20}$ & $2.26\times10^{-20}$ & $8.0\times10^{-21}$ \\
    Dispersion measure DM (pc\,cm$^{-3}$) & 38.775 & 37.049 & 8.090 & 23.918 \\
    DM distance $d_\mathrm{DM}$ (kpc) & 1.70 & 1.90 & 0.44 & 1.56 \\
    Spindown age $\tau_\mathrm{c}$ (Gyr) & 5.5 & 2.8 & 2.6 & 5.6 \\
    Spindown luminosity $L_\mathrm{SD}$ (erg\,s$^{-1}$) & $1.2\times10^{34}$ & $2.2\times10^{34}$ & $6.1\times10^{33}$ & $1.3\times10^{34}$ \\
    Orbital period $P_\mathrm{b}$ (d) & 1.1985 & 53.585 & 1.8601 & 2.1985 \\
    Projected semi-major axis $a \sin i$ (s) & 1.0914 & 27.639 & 2.0426 & 2.1929 \\
    Minimum companion mass $M_\mathrm{c,min}$ (M$_\odot$) & 0.132 & 0.282 & 0.188 & 0.180 \\
    Median companion mass $M_\mathrm{c,med}$ (M$_\odot$) & 0.153 & 0.332 & 0.220 & 0.211 \\
    Predicted companion mass $M_\mathrm{c,pred}$ (M$_\odot$) & 0.188 & 0.299 & 0.198 & 0.201 \\ 
    References & 1 & 2, 3, 4 & 2, 3 & 3, 4, 5 \\
    \hline
    \multicolumn{5}{l}{Derived parameters} \\
    \hline
    Companion right ascension $\alpha_\mathrm{J2000}$ & $\ldots$ & $06^\mathrm{h}14^\mathrm{m}10\fs348(8)$ & $12^\mathrm{h}31^\mathrm{m}11\fs32(1)$ & $20^\mathrm{h}17^\mathrm{m}22\fs706(8)$ \\
    Companion declination $\delta_\mathrm{J2000}$ & $\ldots$ & $-33\degr29\arcmin53\farcs97(9)$ & $-14\degr11\arcmin43\farcs43(16)$ & $+06\degr03\arcmin05\farcs72(18)$ \\
    Offset in right ascension ($\Delta \alpha$) & $\ldots$ & $+0\farcs00(10)$ & $+0\farcs21(18)$ & $+0\farcs02(13)$ \\
    Offset in declination ($\Delta \delta$) & $\ldots$ & $+0\farcs15(9)$ & $+0\farcs24(17)$ & $+0\farcs15(18)$ \\
    $g^\prime$-band magnitude & $>24.7$ & $22.11(1)$ & $25.5(1)$  & $25.30(9)$ \\
    $r^\prime$-band magnitude & $>23.8$ & $21.77(1)$ & $23.91(4)$ & $24.0(2)$  \\
    $i^\prime$-band magnitude & $>23.7$ & $21.69(1)$ & $23.52(7)$ & $24.24(7)$ \\
    Effective temperature $T_\mathrm{eff}$ (K) & $\la$4000 & $6460\pm80$ & $\sim3000$ & $\la2500$ \\
    White dwarf cooling age $\tau_\mathrm{WD}$ (Gyr) & $\ga5$ & 1--1.5 & 5--8 & 9--12 \\
    White dwarf distance $d$ (kpc) & $\ldots$ & 0.54--0.63 & 0.39--0.48 & $<0.5$
    \\
    \hline
  \end{tabular}\\
  References: (1) \cite{dcl+15}; (2) \cite{rrc+11}; (3) \citet{gsl+15};\\
  (4) \citet{abb+15}; (5) \cite{cgj+11}
\end{table*} 

Prior to these blind and targeted surveys the majority of millisecond
pulsar binary companions located in the Galactic field (outside of
globular clusters) were either low-mass (0.1--0.4\,M$_\odot$) or
intermediate mass (0.4--1.1\,M$_\odot$) white dwarfs, with the
exception of two exotic 'black widow' pulsars with very low-mass (a
few 0.01\,M$_\odot$) binary companions \citep{kbjj05}. The new
discoveries have shown us these exotic binary systems are in fact more
common place, and even led to the addition of another class of binary
companions, the 'redback' systems \citep{rob13}. The binary companions
in the latter class have masses in a similar range as the low-mass
white dwarfs (0.1--0.4\,M$_\odot$), but are typically in compact
orbits ($P_\mathrm{b}<1$\,d). Both the black widow and redback systems
are characterized by their compact orbits and eclipses of the radio
signal due to material being ablated from the companion due to
irradiation by the wind from the energetic pulsar. Furthermore, three
of the redback systems have now been observed to transition
back-and-forth between the rotation-powered pulsar state and an
accretion-powered low-mass X-ray binary phase
\citep{asr+09,pfb+13,bph+14}, providing vital clues for the physics
involved in the formation of millisecond pulsar binaries.

Optical observations of binary companions to millisecond pulsars allow
independent measurements of millisecond pulsar and binary
parameters. For the irradiated companions in black widow and redback
systems, modelling of the orbitally modulated light curves constrains
the companion temperatures, the orbital inclination and the
irradiation efficiency of the companion by the pulsar wind (see
\citealt{bkr+13} and references therein). For systems with white dwarf
companions the companion temperature can be determined from
photometry, allowing independent constraints on the white dwarf
cooling age and the distance. The radial velocity curve and the
atmospheric parameters of bright white dwarfs can be measured through
optical spectroscopy, which determine the mass of the white dwarf and
the pulsar (see \citealt{kbjj05} for a review).

Here we report on optical observations of four energetic binary
millisecond pulsars which are seen in $\gamma$-rays by
\textit{Fermi}. The parameters of these binary pulsars are listed in
Table\,\ref{tab:parameters}.  Of the four, PSR\,J0613$-$0200 was
already independently discovered by \citet{lnl+95}, while
PSRs\,J0614$-$3329, J1231$-$1411 and J2017+0603 were discovered in
targeted searches of unassociated sources found by the \textit{Fermi}
LAT instrument \citep{rrc+11,cgj+11}. The orbital parameters of the
systems constrain the companion masses to the range of 0.1 to
0.4\,M$_\odot$, and the absence of radio eclipses suggests that they
are white dwarfs.

\begin{figure*}
  \centering
  \includegraphics[width=0.32\textwidth]{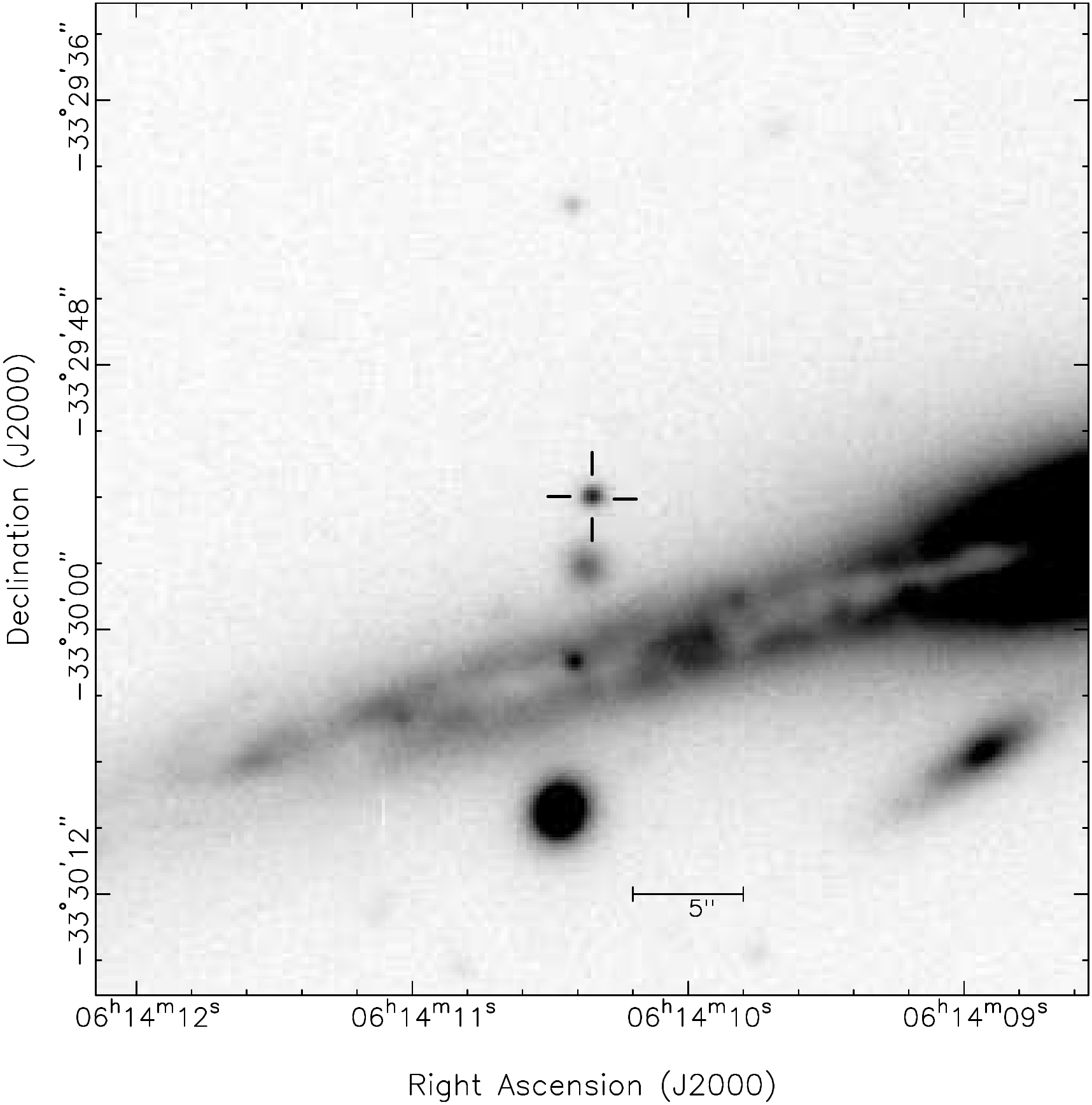}
  \includegraphics[width=0.32\textwidth]{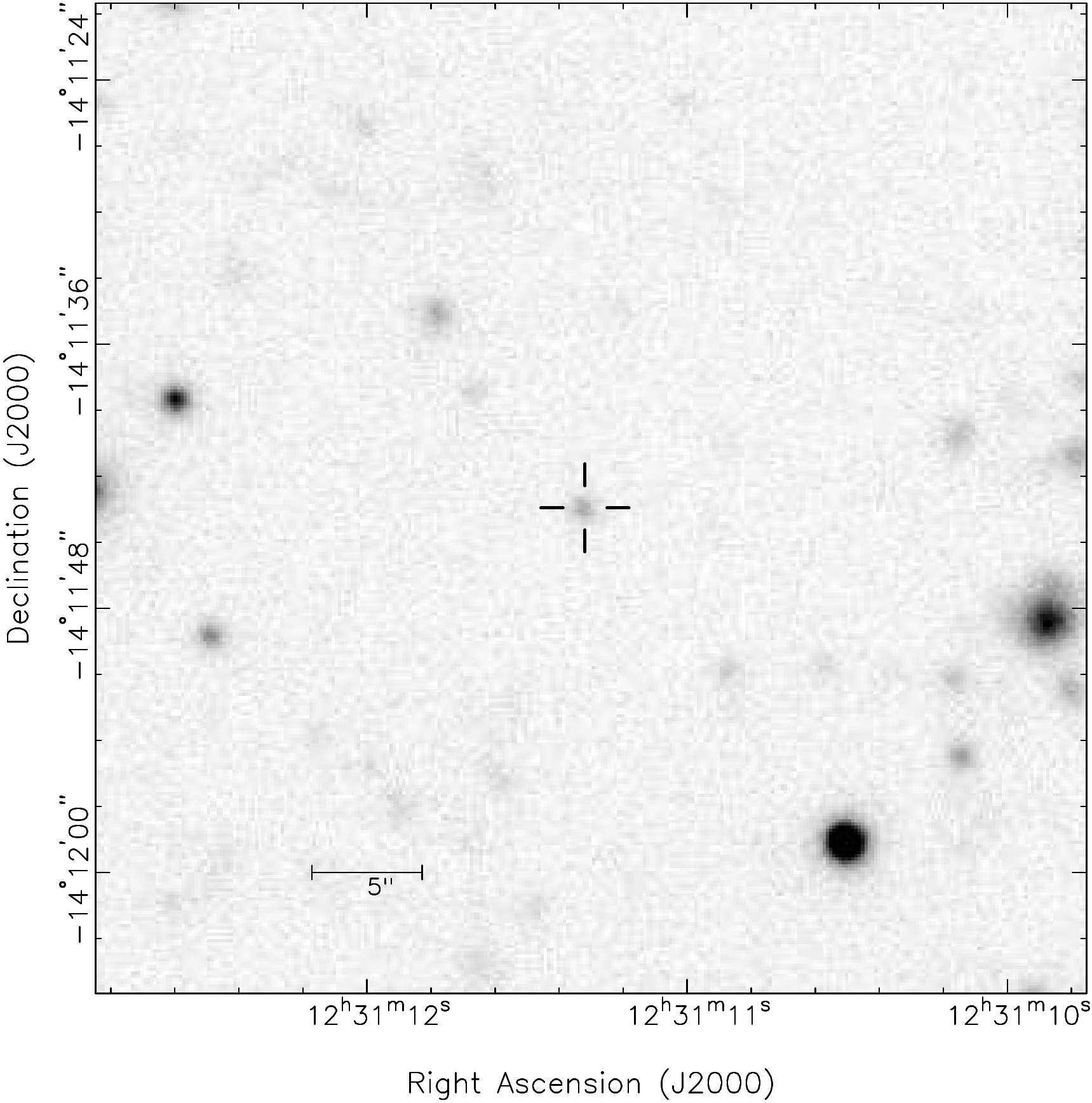}
  \includegraphics[width=0.32\textwidth]{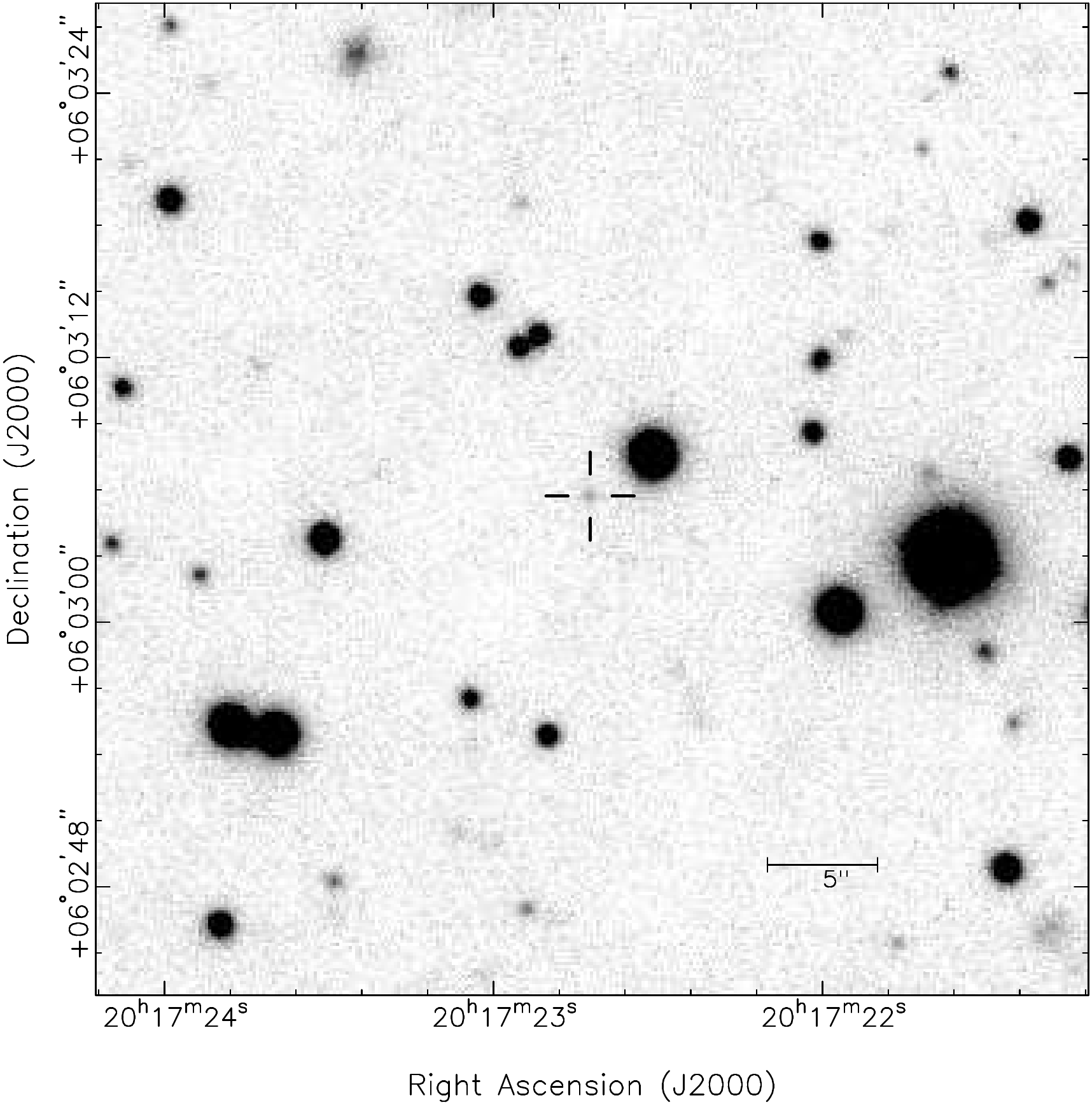}
  \caption{Images of the fields of PSRs\,J0614$-$3329 (left),
    J1231$-$1411 (middle) and J2017+0603 (right). Each image is
    $45\arcsec\times45\arcsec$ in size. The image of J2017+0603 is in
    $i^\prime$, the others are in $r^\prime$. The optical counterpart
    of each pulsar is indicated with $1\arcsec$ long tick marks. }
  \label{fig:charts}
\end{figure*}

\section{Observations and data analysis}
The fields of PSRs\,J0613$-$0200, J0614$-$3329, J1231$-$1411 and
J2017+0603 were observed with the Gemini Multi-Object Spectrograph
(GMOS) on the 8\,m Gemini South telescope in Chile during the fall and
winter of 2010/2011. The GMOS instrument uses three
$2048\times4608$\,pix CCDs and with a pixel scale of
$0\farcs073$\,pix$^{-1}$ they provide a vignetted field-of-view of
$5\farcm5\times5\farcm5$. For these observations only data from the
center CCD was used providing a field-of-view of
$2\farcm5\times5\farcm5$. The CCD was read using $2\times2$ binning
with a scale of $0\farcs146$\,pix$^{-1}$. Sets of dithered images were
obtained in the Sloan $g^\prime$, $r^\prime$ and $i^\prime$ filters
\citep{fig+96} with 4\,min exposures for PSRs\,J0614$-$3329,
J1231$-$1411 and J2017+0603. Shorter exposures of 30\,sec, 1\,min and
2\,min were obtained for PSR\,J0613$-$0200 to prevent a nearby star
\citep{kbjj05} from saturating.  For the astrometric and photometric
calibration, short, 5 to 30\,s exposures using all three filters were
obtained for all three sources under photometric conditions. Flatfield
frames were obtained during twilight through the standard Gemini
calibration plan. The seeing during the observations varied between
$0\farcs7$ to $1\farcs1$.

Long-slit spectroscopy of the counterpart to PSR\,J0614$-$3329 was
obtained with FORS2 at the ESO VLT. Two exposures with integration
times of 3600\,s and 3000\,s were obtained on 2011 December 20 and 21
using the 1200B grism covering the wavelength range from 3730 to
5190\,\AA. A $1\arcsec$ slit was used with $2\times2$ binning,
yielding a resolution of 2.8\,\AA, sampled at
0.71\,\AA\,pix$^{-1}$. The seeing during these observations varied
between $0\farcs7$ and $1\farcs1$. In both observations the slit was
placed such to include the three point-like sources South of the
companion, as well as some of the emission from a background
galaxy (see Fig.\,\ref{fig:charts}).

\subsection{Astrometry}
Stars on the short (5 to 30\,s) $r^\prime$ images of each field were
matched against stars in the 2MASS (\citealt{scs+06}; for
PSRs\,J0613$-$0200, J0614$-$3329 and J2017+0603) or USNO-B1
(\citealt{mlc+03}; for PSR\,J1231$-$1411) catalogs for astrometric
calibration. Fitting for a zero-point offset and a 4 parameter
transfer matrix, an astrometric solution for each pulsar was
determined having rms residuals ranging from $0\farcs1$ to $0\farcs2$
in each coordinate. For each pulsar the astrometric solution was
transferred to the median combined images using a few dozen stars
common to both images with considerably smaller rms residuals of
$0\farcs02$ in both coordinates.

The calibrated images were searched for optical counterparts
consistent with the pulsar timing positions. In the case of
PSR\,J0613$-$0200 we use the timing position and proper motion of
\citet{dcl+15}, which yields
$\alpha_\mathrm{J2000}=06^\mathrm{h}13^\mathrm{m}43\fs9760$ and
$\delta_\mathrm{J2000}=-02\degr00\arcmin47\farcs241$ for the epoch of
the GMOS observations. The uncertainty on the position is
negligible. \citet{rrc+11} provides timing positions for
PSRs\,J0614$-$3329 and J1231$-$1411, of which the latter has a
somewhat uncertain proper motion which places the pulsar at
$\alpha_\mathrm{J2000}=12^\mathrm{h}31^\mathrm{m}11\fs305(2)$ and
$\delta_\mathrm{J2000}=-14\degr11\arcmin43\farcs67(6)$ at the time of
the optical observations. Using the more accurate proper motion by
\citet{gsl+15} yields a consistent position. For PSR\,J2017+0603 we
use the position from \citet{cgj+11}.

\begin{figure}
  \includegraphics[width=0.33\textwidth]{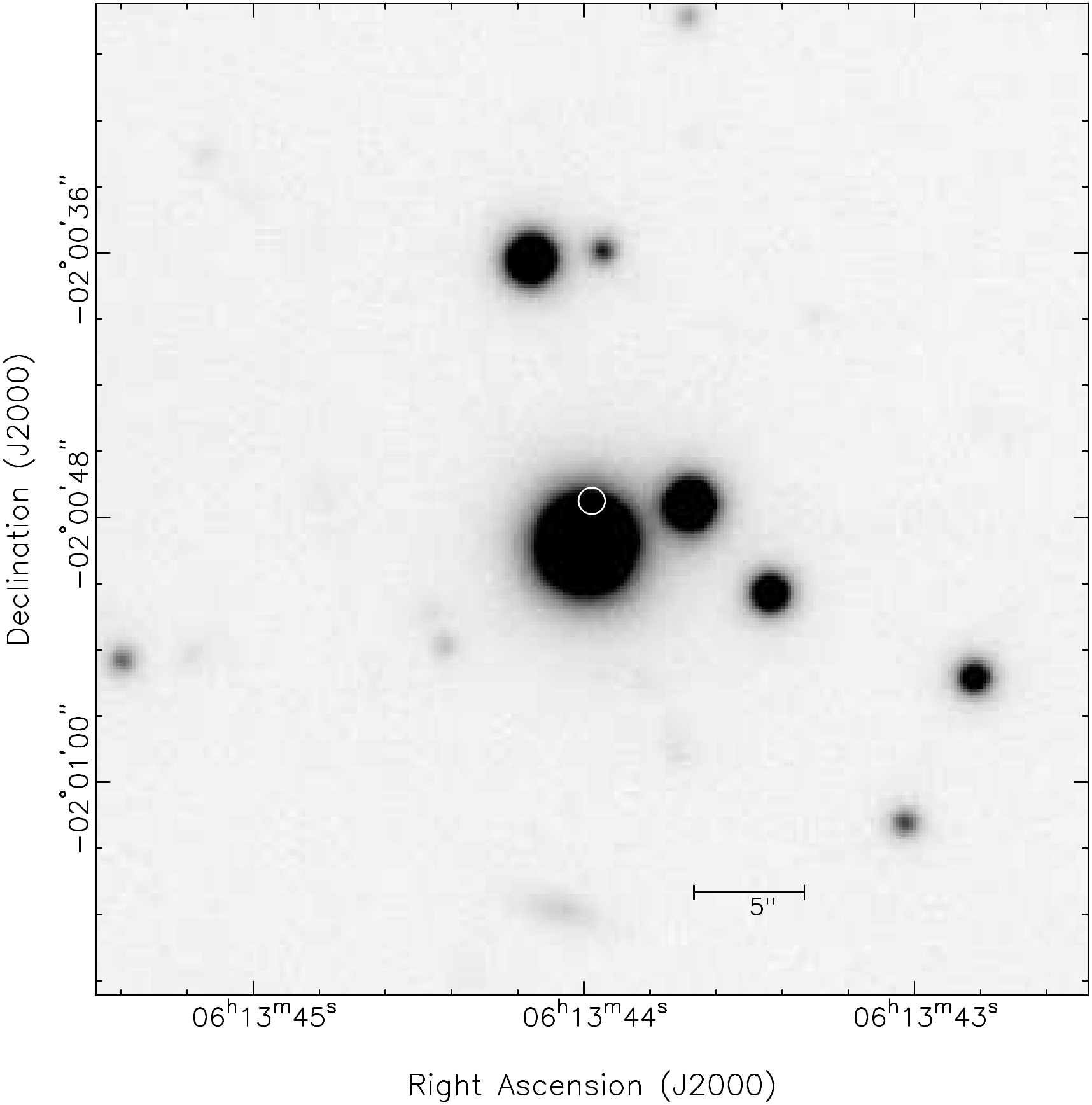}
  \includegraphics[width=0.14\textwidth]{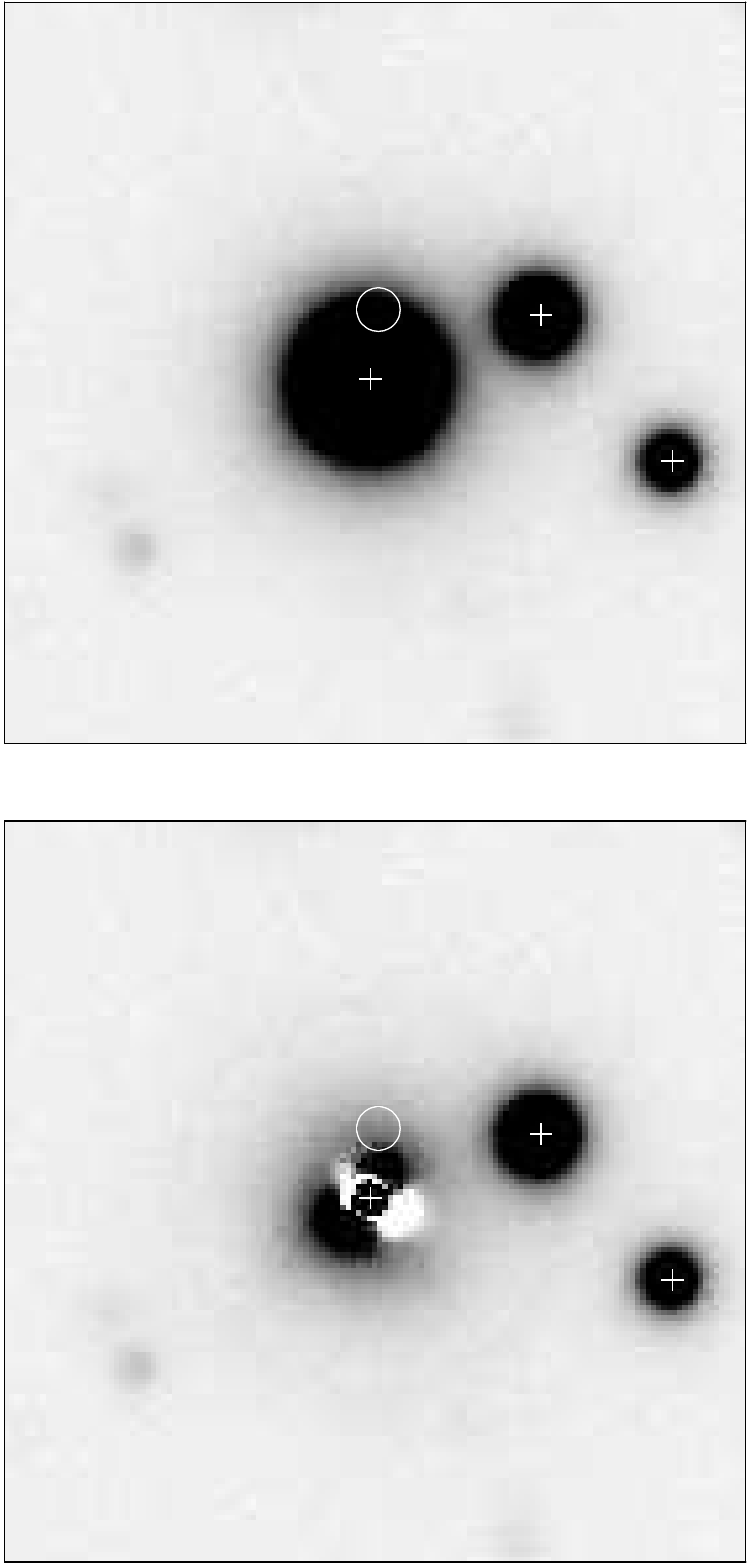}
  \caption{Images of the field containing PSR\,J0613$-$0200. The
    left-hand panel shows a $45\arcsec\times45\arcsec$ subsection of
    the averaged $r^\prime$-band image. The radio timing position of
    PSR\,J0613$-$0200 is indicated with the $0\farcs6$ (99\%
    confidence) error circle. The two right-hand panels show a
    $20\arcsec\times20\arcsec$ subsection of the same image, where in
    the bottom panel the stellar profile of the nearby star is
    removed. No counterpart is visible within the error circle.}
  \label{fig:j0613chart}
\end{figure}

Optical counterparts are present in the 95\% confidence error regions
of the timing positions of J0614$-$3329, J1231$-$1411 and J2017+0603
(see Fig.\,\ref{fig:charts}). The celestial position and observed
$g^\prime r^\prime i^\prime$ magnitudes of the counterparts are given
in Table\,\ref{tab:parameters}. In the case of PSR\,J0613$-$0200 the
pulsar position is offset by $1\farcs9$ from a nearby unassociated
star, having $r^\prime=15.6$. No counterpart is present in the
$0\farcs60$ (99\% confidence) error circle, see
Fig.\,\ref{fig:j0613chart}.

\subsection{Photometry}
The images were reduced using standard methods, where all images were
corrected for bias from the overscan regions and flatfielded using the
twilight flats. The images taken with the $i^\prime$ filter suffered
from fringing. To correct for this, a fringe frame was created by
median combining all 4\,min $i^\prime$ images of PSR\,J2017+0603 and
PSR\,J0614$-$3329 such that it contained only the contributions of the
sky and the fringe variations. The level of the sky was estimated and
subtracted, and the remaining fringe variations were scaled and used
to remove the fringes from the individual $i^\prime$ images. Finally,
the dithered images in each filter were aligned using integer pixel
offsets and median combined to increase the overall signal-to-noise.

The DAOPHOT\,II package \citep{ste87} was used to determine
instrumental magnitudes of stars on the images through
point-spread-function (PSF) fitting. Aperture photometry of several
bright stars was used to determine aperture corrections. For the
photometric calibration we determined and compared instrumental
magnitudes of stars in three standard fields (060000$-$300000, T\,Phe
and PG\,0942$-$029), observed in October and November 2010 as part of
the standard GMOS calibration plan, against calibrated $g^\prime
r^\prime i^\prime$ magnitudes from the updated
catalog\footnote{\url{http://www-star.fnal.gov/Southern\_ugriz/www/Fieldindex.html}}
of \citet{stk+02}. Fitting for zero-point and color-coefficient,
assuming standard GMOS extinction coefficients of 0.18, 0.10 and
0.08\,mag per airmass for $g^\prime$, $r^\prime$ and $i^\prime$,
respectively, we obtained rms residuals of 0.05\,mag in $g^\prime$,
0.02\,mag in $r^\prime$ and 0.04\,mag in $i^\prime$.

Limits on the magnitude of a counterpart to PSR\,J0613$-$0200, located
in the wings of the nearby star, were computed by performing
photometry on copies of the original image with an artificial star
inserted on the position of the pulsar. By varying the magnitude of
the artificial star, $5\sigma$ detection limits were derived. These are
listed in Table\,\ref{tab:parameters}.

Finally, we checked for variability of the detected counterparts by
comparing instrumental magnitudes obtained from the individual
frames. For the counterpart to PSR\,J0614$-$3329 we do not find
significant variability down to rms scatter of less than 0.05\,mag in
all three filters. This scatter is comparable to the scatter seen in
stars of similar brightness. The counterparts to PSRs\,J1231$-$1411
and J2017+0603 are considerably fainter and not detected in each
individual frame. As such, we set a conservative limit on the
variability of the counterparts in these two systems at less than
0.3\,mag in all three filters.

\subsection{Spectroscopy}
The spectra were corrected for bias by subtracting a master bias frame
and flatfielded using lamp-flats. The spectrum of the companion was
optimally extracted using a variation on the algorithms by
\citet{hor86} and \citet{hyn02} to separate the flux from the
companion from that of the nearby object by modelling their wavelength
dependent spatial profiles. The wavelength scale was calibrated
against arc lamp exposures taken during daytime. For the flux
calibration we used an observation of the spectrophotometric standard
GD\,108 taken as part of the normal VLT calibration program. This
calibration is approximate, as no correction for slit losses was
applied. Based on the width of the slit and the seeing during the
PSR\,J0614$-$3329 observations, at least 95\% of the flux is accounted
for.

\begin{figure}
  \centering
  \includegraphics[width=\columnwidth]{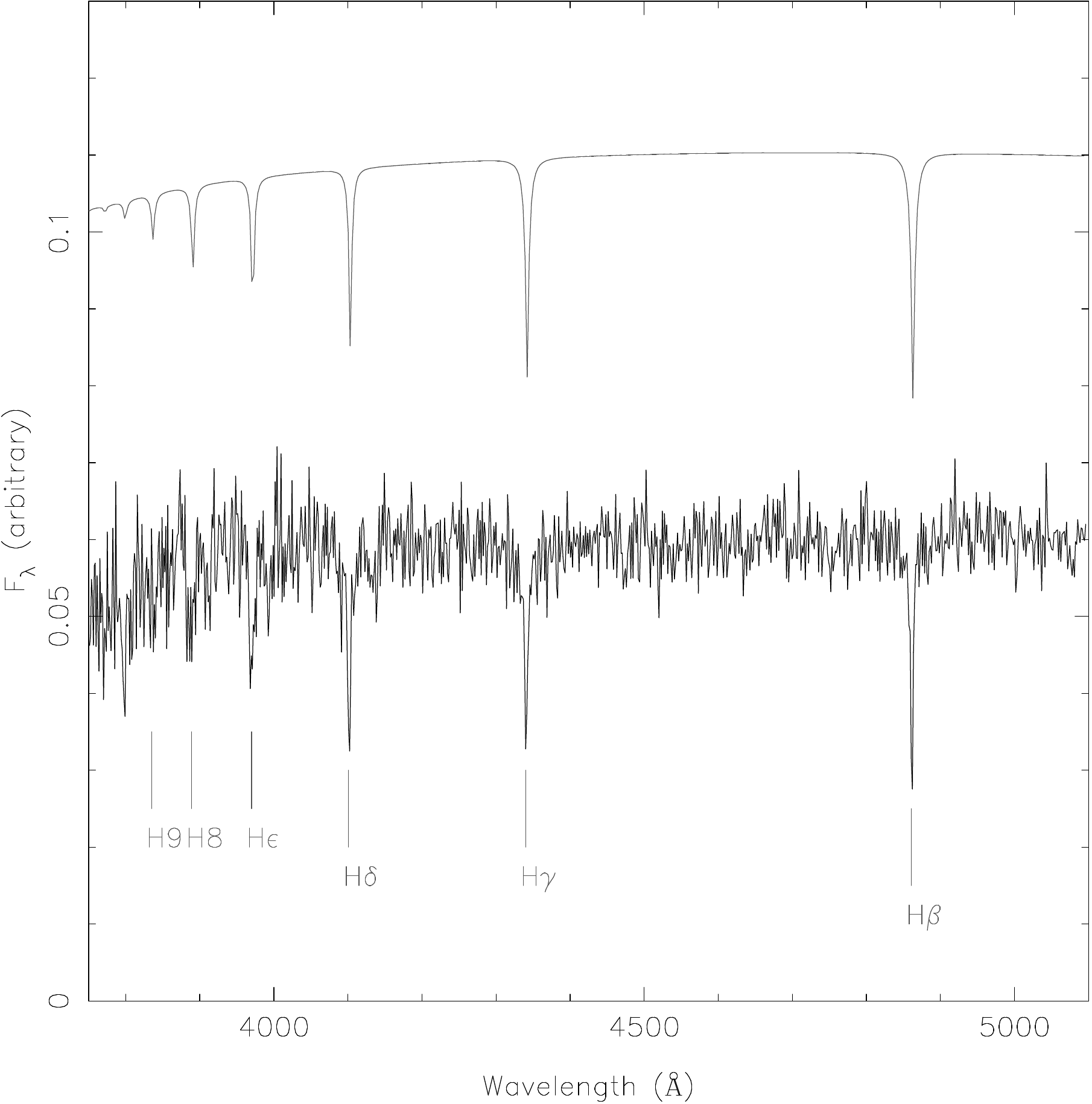}
  \caption{The spectrum of the counterpart to PSR\,J0614$-$3329. The
    lower curve shows the average of the two VLT spectra, shifted to
    zero velocity. Hydrogen lines in the Balmer series are seen from
    H$\beta$ upto H9, consistent with a DA type white dwarf. The top
    curve is a white dwarf atmosphere model with
    $T_\mathrm{eff}=6500$\,K and $\log g=7.0$\,cgs, broadened by a
    truncated Gaussian representing the response of the slit and
    seeing and scaled to fit the observed spectrum. This model lies
    closest in the grid of atmosphere models used to the best fit
    values of $T_\mathrm{eff}=6460\pm80$\,K and $\log
    g=7.0\pm0.2$\,cgs. Note that the model spectrum has been shifted
    upwards by 0.05 units. }
  \label{fig:j0614spectrum}
\end{figure}

\section{Results}
We have found optical counterparts to PSR\,J0614$-$3329,
PSR\,J1231$-$1411 and PSR\,J2017+0603. Given the density of stars in
these fields, we estimate that the probability of finding a star in
the error circle of these objects by chance is only a few percent. As
such, we treat the optical counterparts to these millisecond pulsars
as their binary companions in the remainder of the paper.

In the case of PSR\,J0614$-$3329, the optical spectrum of the
counterpart (Fig.\,\ref{fig:j0614spectrum}) shows the H$\beta$ to H9
hydrogen Balmer lines, classifying it as a DA type white dwarf,
confirming it is the binary companion. The effective temperature and
the surface gravity can be constrained by comparison of the observed
spectrum with a grid of theoretical Hydrogen atmosphere models
appropriate for DA type white dwarfs. This grid, an update of that
presented in \citet{koe10}, spans temperatures of
$T_\mathrm{eff}=6000$ to 7500\,K and surface gravities of $\log g=6.0$
to 8.5\,cgs, with steps of 250\,K and 0.25\,dex. Using the fitting
method described in \citet{bkkv06}, the atmospheric parameters are
constrained to $T_\mathrm{eff}=6460\pm80$\,K and $\log
g=7.0\pm0.2$\,cgs.

\begin{figure*}
  \includegraphics[width=0.8\textwidth]{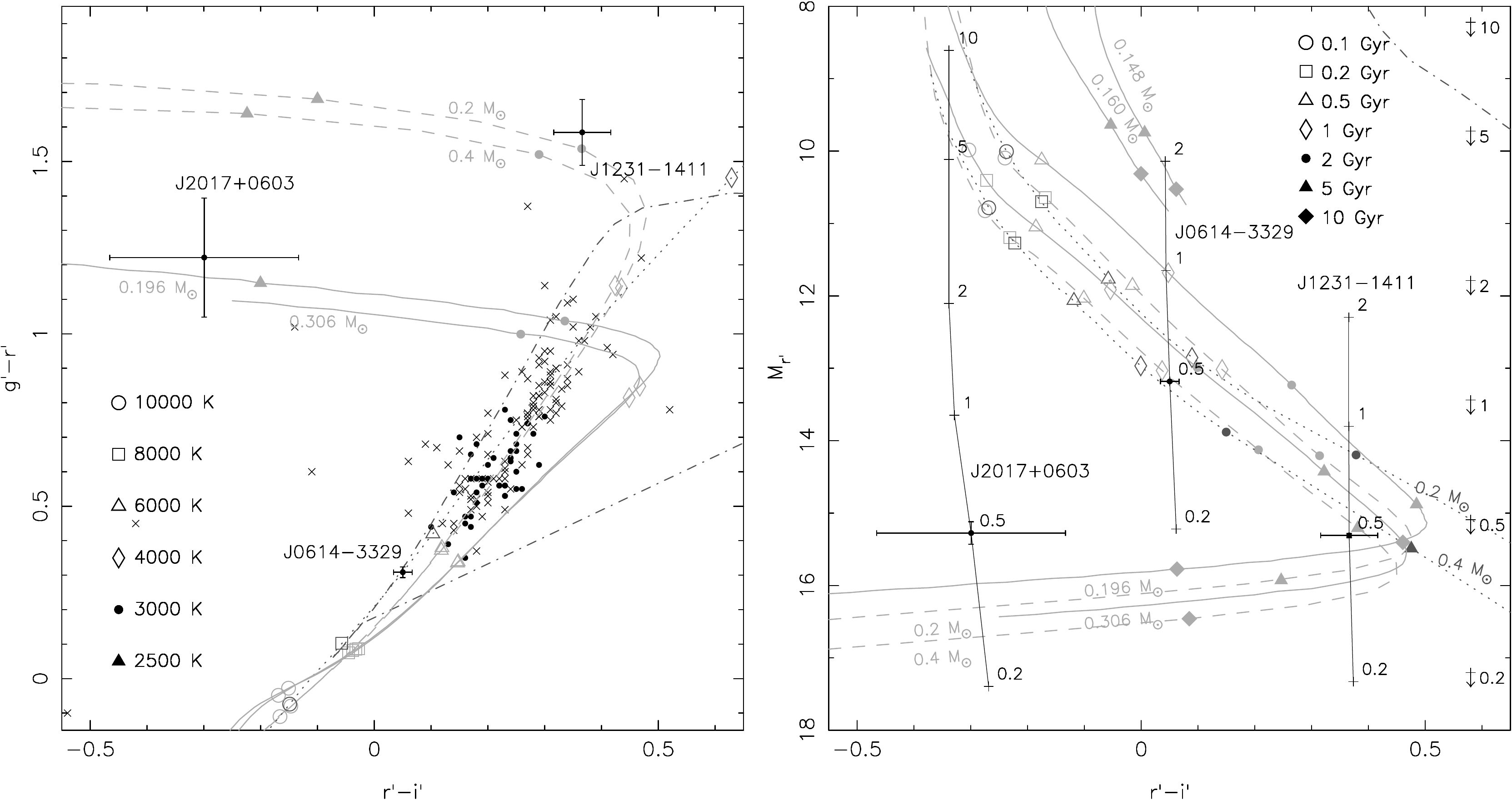}
   \caption{Color-color \textit{(left)} and color-magnitude
     \textit{(right)} diagrams showing various Helium-core white dwarf
     cooling tracks. The \citet{rsab02} models for specific white
     dwarf masses are plotted as solid lines, while the Bergeron
     models as described in \citet{hb06,ks06,tbg11} and \citet{bwd+11}
     are dashed for white dwarfs with a Hydrogen atmosphere (DA white
     dwarfs), and dotted for those with Helium atmospheres (DB white
     dwarfs). The Johnson-Cousins $BVRI$ magnitudes from the
     \citet{rsab02} model were converted to SDSS $g^\prime r^\prime
     i^\prime$ using the transformations of \citet{stk+02}. The mass
     of each model is indicated. Models with masses below about
     0.2\,M$_\odot$ have thick Hydrogen envelopes and cool slower than
     models with higher masses. Effective temperatures or cooling ages
     are indicated along the tracks. The dash-dotted line is a 1\,Gyr,
     solar metallicity isochrone from \citet{ggoc04} depicting the
     location of non-degenerate main-sequence stars. In the
     color-color diagram the colors of ultra-cool white dwarfs from
     the SDSS \citep{klt+10} are plotted. White dwarfs with
     spectroscopically determined Hydrogen atmospheres (DA) are
     plotted with dots; those with non-Hydrogen atmospheres are
     plotted with crosses (mostly DC type showing no spectral
     features). The observed $g^\prime-r^\prime$ and
     $r^\prime-i^\prime$ colors for PSRs\,J0614$-$3329, J1231$-$1411
     and J2017+0603 are plotted in the color-color diagram, while
     $r^\prime-i^\prime$ colors and absolute $r^\prime$ magnitudes
     ($M_{r^\prime}$) are shown in the color-magnitude diagram. These
     absolute magnitudes are plotted for various distances indicated
     by the numbers along the track (distance in kiloparsecs) and
     corrected for absorption and reddening using the predictions from
     the \citet{dcl03} model, using extinction coefficients from
     \citet{sfd98}. Finally, limits on the absolute $r^\prime$
     magnitude of PSR\,J0613$-$0200 are indicated by arrows on the
     right-hand side of the color-magnitude diagram. Here various
     distances (in kiloparsecs) are again indicated.}
   \label{fig:cmd}
\end{figure*}

Interpolating between the discrete mass models by \citet{pach07},
these atmospheric parameters correspond to a white dwarf mass of
$M_\mathrm{WD}=0.24\pm0.04$\,M$_\odot$. The minimum companion mass for
an edge on orbit and a 1.4\,M$_\odot$ pulsar is 0.282\,M$_\odot$
($M_\mathrm{c,min}$ in Table\,\ref{tab:parameters}), while the
$P_\mathrm{b}$-$M_\mathrm{c}$ relation between orbital period and
companion mass, as predicted by binary evolution, predicts
$M_\mathrm{c,pred}=0.299$\,M$_\odot$ \citep{lrp+11}. As the inferred
mass is lower than that, this may indicate that we observe the system
edge on or that the pulsar may be less massive than the assumed
1.4\,M$_\odot$. However, we consider it more likely that the low
signal-to-noise of the observed spectrum and degeneracies between the
shape of the Balmer lines and the broadening by the slit and seeing
leads to unmodelled systematic errors underestimating the surface
gravity.

In Fig.\,\ref{fig:cmd} we compare the broadband magnitudes of the
millisecond pulsar companions against predictions from white dwarf
cooling models by \citet{rsab02} and those by Bergeron et
al.\footnote{\url{http://www.astro.umontreal.ca/~bergeron/CoolingModels/}}
as described in \citet{hb06,ks06,tbg11} and \citet{bwd+11}. Here we
used the model for Galactic absorption by \citet{dcl03}, which
predicts $A_V$ as a function of distance for specific lines of
sight. The extinction coefficients by \citet{sfd98} were used to
transform $A_V$ into $A_{g^\prime}$, $A_{r^\prime}$ and
$A_{i^\prime}$. Because of the high Galactic latitude of the pulsars,
the absorption is small; PSR\,J0613$-$0200 has the largest maximum
absorption of $A_{V,\mathrm{max}}=0.7$\,mag, while PSRs\,J0614$-$3329,
J1231$-$1141 and J2017+0603 have 0.15, 0.13 and 0.46\,mag,
respectively.

We find that for PSR\,J0614$-$3329 the comparison of the broadband
colors with cooling models yields a temperature that is consistent
with the spectroscopic determination. At these temperatures, the
Bergeron and \citet{rsab02} white dwarf cooling models yield distances
in the range of $d=540$ to 630\,pc and white dwarf cooling ages
between 1 and 1.5\,Gyr.

In the case of PSR\,J1231$-$1411 and PSR\,J2017+0603, the observed
colors and magnitudes of the companions are consistent with those
expected for cool white dwarfs. Main-sequence companions such as those
observed in the ``redback'' systems can be ruled out; the colors
deviate significantly from those of main-sequence stars. For
comparison, a 0.2\,M$_\odot$ main-sequence star has
$M_{r^\prime}=11.49$, $g^\prime-i^\prime=1.42$ and
$r^\prime-i^\prime=0.99$ \citep{ggoc04}. Furthermore, ``redback''
systems have severely bloated companions, and the companion to the
``redback'' PSR\,J1023+0038 has $r^\prime=17.43$, which, at a distance
of $d=1.37$\,kpc \citep{dab+12}, would translate to
$M_{r^\prime}=6.75$. If the companions to PSRs\,J1231$-$1411 and
J2017+0603 had that intrinsic brightness they would be at distances of
$d=28$\,kpc, easily ruled out by the dispersion measure distance
estimates.
\vspace{4mm}

Instead, we find that for PSR\,J1231$-$1411 and PSR\,J2017+0603, the
$g^\prime-r^\prime$ and $r^\prime-i^\prime$ colors of the companions
are located in the part of the color-color diagram where the dominant
source of opacity changes from bound-free absorption of $H^-$ at high
temperatures to collision-induced absorption of H$_2$ at lower
temperatures \citep{lcs91,sbl+94,han98}. The Bergeron and
\citet{rsab02} models diverge here as the latter do not take into
account the effect of Ly\,$\alpha$ absorption on the opacity at blue
wavelengths \citep{ks06}. Furthermore, depending on the atmospheric
composition of the white dwarf (the He/H abundance ratio) the cooling
tracks can level out at $g^\prime-r^\prime$ values ranging from 0.7 to
1.6 (\citealt{gbd12}, see also Fig.\,7 of \citealt{pgm+12}). The
updated models by \citet{rak11} do take into account the absorption by
Ly\,$\alpha$, but unfortunately only cover white dwarf masses between
0.5 and 0.9\,M$_\odot$, which is above the mass range we expect for
the systems studied in this paper.

Despite the uncertainty in the Helium abundance and the predicted
$g^\prime-r^\prime$ color, the Bergeron and \citet{rsab02} cooling
models predict comparable temperatures and absolute magnitudes for a
given $r^\prime-i^\prime$ color. Taking the predictions at face value,
we find that the companions to PSR\,J1231$-$1411 and PSR\,J2017+0603
must be cool ($T_\mathrm{eff}\la3000$\,K), old
($\tau_\mathrm{WD}\ga7$\,Gyr) and nearby ($d\la0.5$\,kpc).

For PSR\,J0613$-$0200, \citep{dcl+15} and \citet{abb+15} report a
parallax of $\pi=1.25\pm0.13$\,mas and $\pi=0.91\pm0.15$\,mas,
respectively. These translate to Lutz-Kelker bias \citep{lk73,vlm10}
corrected distance estimates of $d=0.78\pm0.08$\,kpc and
$d=0.98\pm0.16$\,kpc. At these distances the non-detection of the white
dwarf companion to PSR\,J0613$-$0200 sets a lower limit on the
absolute $r^\prime$ magnitude of $M_{r^\prime}>13.9$. This limit
excludes the thick Hydrogen atmosphere models by \citet{rsab02} (those
with masses of 0.148, 0.160 and 0.169\,M$_\odot$), which still have
$M_V\sim M_{r^\prime}\la11$ at ages of 15\,Gyr. As such, the
observations suggest that the companion must have a thin Hydrogen
atmosphere and is expected to have temperatures below 5000\,K and have
cooling ages in excess of 4\,Gyr.

\section{Discussion and conclusions}
We have discovered the binary companions to radio millisecond pulsars
J0614$-$3329, J1231$-$1411 and J2017+0603, and set limits on the
brightness of the companion to PSR\,J0613$-$0200.

The spectroscopic observations of the companion to PSR\,J0614$-$3329
confirm that the companion is a DA type white dwarf and constrain the
temperature to $T_\mathrm{eff}=6460\pm80$\,K. The surface gravity of
$\log g=7.0\pm0.2$\,cgs and hence white dwarf mass of
$M_\mathrm{WD}=0.24\pm0.04$\,M$_\odot$ are low but consistent with expectations
based on the orbital parameters of the binary. Comparing the broadband
photometry against white dwarf cooling models yields a distance which
is a factor 3 to 4 closer than the $d_\mathrm{DM}=1.9$\,kpc predicted
by the NE2001 model \citep{cl02} based on the observed dispersion
measure. \citet{rrc+11} already noted that the NE2001 model was over
estimating the distance to PSR\,J0614$-$3329 due to its location at
the edge of the Gum nebula. This reduction in the pulsar distance
equally reduces the very high implied $\gamma$-ray efficiency of
$\eta=215$\% \citep{rrc+11,jvh+14} to more appropriate values of order
17 to 24\%.

The distance of the white dwarf companion to PSR\,J1231$-$1411 is
consistent with the distance estimated from the dispersion measure of
the pulsar. In the case of PSR\,J2017+0603 the white dwarf cooling
models place the system at a distance of less than 0.5\,kpc. This is
considerably closer than the dispersion measure distance of
$d_\mathrm{DM}=1.56$\,kpc, and consistent with the \citet{lk73} bias
correct parallax distance of $0.6\pm0.2$\,kpc measured from
pulsar timing \citep{gsl+15}. The distance determined from the white
dwarf photometry would reduce the $\gamma$-ray efficiency of
PSR\,J2017+0603 to values below 8\%.

We find that the companions to PSRs\,J0613$-$0200, J1231$-$1411 and
J2017+0603 are old and cool. They are similar in temperature or cooler
than the best studied low-mass pulsar companion, that of
PSR\,J0437$-$4715, for which \citet{dkp+12} derive a temperature of
$T_\mathrm{eff}=3950\pm150$\,K. To cool to these low temperatures the
white dwarfs must have thin Hydrogen envelopes to prevent residual
Hydrogen burning keeping the white dwarf hot
(e.g.\ \citealt{hp98a}). The thickness of the Hydrogen envelope can be
reduced by Hydrogen shell flashes. Evolutionary models indicate that
Hydrogen shell flashes only occur for white dwarfs with masses above
about 0.2\,M$_\odot$, though it is not clear if this is a sharp
threshold, leading to a dichotomy in white dwarf cooling properties
\citep{ashp96,dsbh98,asb01,amc13}, or a smooth transition with white
dwarf mass \citep{itla14}.

\begin{figure}
  \centering
  \includegraphics[width=\columnwidth]{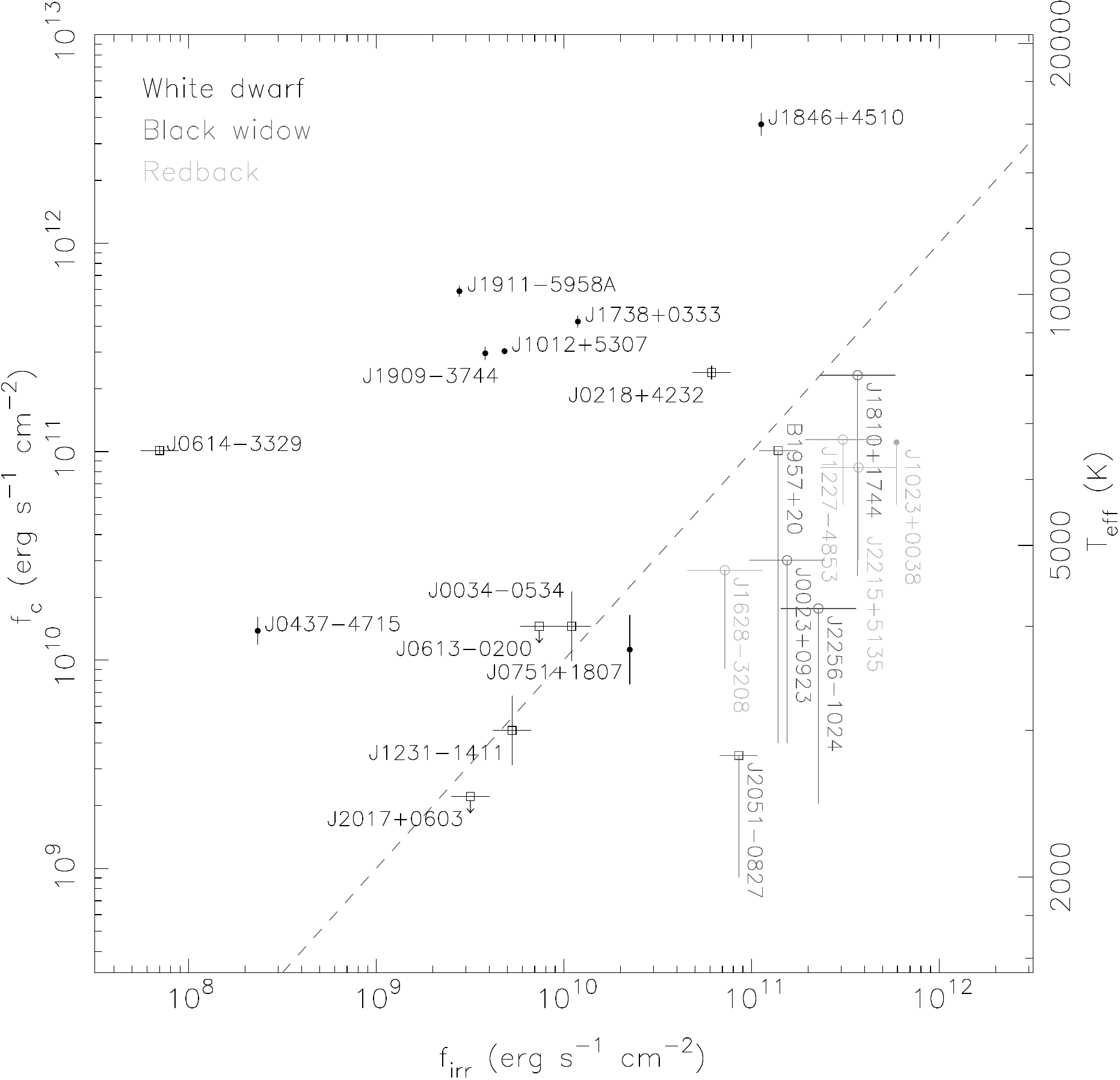}
  \caption{A comparison between the flux of the companion
    $f_\mathrm{c}$ and the flux of the pulsar wind irradiating the
    companion $f_\mathrm{irr}$ for pulsar binaries where accurate
    companion temperatures are available. The spindown luminosity has
    used spin period derivatives corrected for the apparent
    acceleration due to proper motion (the \citealt{shk70} effect)
    where available. The binary separation $a$ is taken for the median
    companion mass (inclination $i=60\degr$) if no mass measurements
    are available. In the case of known proper motions and binary
    masses, the points are indicated with black dots. If only proper
    motions are known, open circles are used and a 10\% uncertainty on
    the spindown luminosity is assumed. If neither proper motion nor
    binary masses are known, a 20\% uncertainty is assumed. The
    companion types are as indicated, while the dashed diagonal line
    depicts $f_\mathrm{c}=f_\mathrm{irr}$. Companion temperatures and
    pulsar parameters are from
    \citet{kbk96,lfc96,cgk98,bkk03,bkk06,bkkv06,dkp+12,jbk+03,akk+12,kbk+13,bkr+13,ta05,sklk99,lht14,mcm+14}
    and the ATNF pulsar catalogue \citep{mhth05}.}
  \label{fig:irradiation}
\end{figure}

Besides Hydrogen flashes, irradiation by the pulsar may result in the
removal of the Hydrogen envelope when the companion detaches from the
Roche lobe. This scenario has been proposed for PSR\,J0751+1807 by
\citet{esa01} to explain the low temperature of its white dwarf
companion. For this pulsar, Shapiro delay measurements yield a white
dwarf mass of 0.16\,M$_\odot$ \citep{nsk08,dcl+15}, below the mass
threshold where no Hydrogen flashes are expected to occur, yet
observations of the companion by \citet{bkk06} show that it is cool
($T_\mathrm{eff}\approx4000$\,K) and faint, ruling out the presence of
a thick Hydrogen envelope.

For the companion to PSR\,J0613$-$0200, and perhaps the companions to
PSRs\,J1231$-$1411 and J2017+0603 irradiation by the pulsar wind may
be, or may have been, important. In Figure\,\ref{fig:irradiation} we
compare the companion flux ($f_\mathrm{c}=\sigma T_\mathrm{eff}^4$) to
the flux of the pulsar wind irradiating the companion surface for all
known optical companions ($f_\mathrm{irr}=L_\mathrm{SD}/4\pi a^2$,
where $L_\mathrm{SD}\propto\dot{P}/P^3$ is the pulsar spindown
luminosity and $a$ is orbital separation). We find that these three
systems, as well as PSR\,J0751+1807, and possibly J0034$-$0534, are
located in the region where $f_\mathrm{c}\approx
f_\mathrm{irr}$. Systems which are clearly irradiated, like the 'black
widow' and 'redback' pulsars, fall in the region where
$f_\mathrm{c}<f_\mathrm{irr}$, while systems with spectroscopically
constrained white dwarf temperatures have
$f_\mathrm{c}>f_\mathrm{irr}$.

Theoretical modelling of the binary evolution of short orbital period
neutron stars has revealed that irradiation plays a major role in the
formation of the 'black widow' and 'redback' systems
\citep{prp03,ccth13,bvh14,bvh15b,bvh15a}. After the companion
decouples from the Roche lobe, the pulsar mechanism switches on and
starts irradiating the companion, leading to mass loss through
evaporation. The efficiency of the irradiation and evaporation
determines the mass loss. These models allow for the possibility
suggested by \citet{esa01}; that (part of) the Hydrogen envelope is
lost due to irradiation after the companion detaches from the Roche
lobe.

Though the white dwarf companions to PSRs\,J0613$-$0200, J1231$-$1411
and J2017+0603 may have had their Hydrogen envelopes reduced through
Hydrogen flashes, we can not rule out the possibility of mass loss due
to irradiation. A search for photometric variability for
PSRs\,J1231$-$1411 and J2017+0603 may constrain the effects of
irradiation in these objects.

\section*{Acknowledgments}
We thank the anonymous referee for useful suggestions, and thank
Marten van Kerkwijk, Thomas Tauris and Alina Istrate for helpful
discussions. CGB acknowledges the hospitality of the International
Space Science Institute (ISSI) in Bern, Switzerland. Based on
observations made with ESO Telescopes at the La Silla Paranal
Observatory under programme ID 088.D-0138 and on observations obtained
at the Gemini Observatory (Program ID GS-2010B-Q-56), which is
operated by the Association of Universities for Research in Astronomy,
Inc., under a cooperative agreement with the NSF on behalf of the
Gemini partnership: the National Science Foundation (United States),
the National Research Council (Canada), CONICYT (Chile), the
Australian Research Council (Australia), Minist\'{e}rio da
Ci\^{e}ncia, Tecnologia e Inova\c{c}\~{a}o (Brazil) and Ministerio de
Ciencia, Tecnolog\'{i}a e Innovaci\'{o}n Productiva (Argentina). CGB
acknowledges support from the European Research Council under the
European Union's Seventh Framework Programme (FP/2007-2013) / ERC
Grant Agreement nr. 337062 (DRAGNET; PI Jason Hessels).

\bibliographystyle{mnras}
%\bibliography{references}

\bsp
\label{lastpage}
\end{document}